\documentclass[10pt]{iopart}
\usepackage{graphicx}
\usepackage[authoryear]{natbib}
\RequirePackage{color}

\begin{document}
\title[Pant et al; A new class of viable and exact solutions of EFE's with.....]{A new class of viable and exact solutions of EFE's with Karmarkar conditions: An application to cold star modeling}
\author{Neeraj Pant$^1$, Megandhren Govender$^2$, \& Satyanarayana Gedela$^3$} 
\address{$^{1,3}$Department of Mathematics, National Defence Academy, Khadakwasla, Pune-411023, India}
\address{$^{2}$ Department of Mathematics, Faculty of Applied Sciences, Durban University of Technology, Durban, South Africa}
\address{$^{3}$ Department of Mathematics, SSJ Campus, Kumaun University, Almora-263601, India.}
\ead{neeraj.pant@yahoo.com$^1$, megandhreng@dut.ac.za$^2$, satya235@gmail.com$^3$}
\vspace{5pt}
\begin{indented}
\item[]August 2020
\end{indented}
\begin{abstract}
In this work we present a theoretical framework within Einstein's classical general relativity which models stellar compact objects such as PSR J1614-2230 and SAX J1808.4-3658. The Einstein field equations are solved by assuming that the interior of the compact object is described by a class I spacetime. The so-called Karmarkar condition arising from this requirement is integrated to reduce the gravitational behaviour to a single generating function. By appealing to physics we adopt a form for the gravitational potential which is sufficiently robust to accurately describe compact objects. Our model satisfies all the requirements for physically realistic stellar structures.
\end{abstract}
\vspace{3pt}
\noindent{\it Keywords}: Compact star, anisotropy, embedding class, Einstein field equations, adiabatic index.\\

\vspace{2pt}
\submitto{\RAA}
\ioptwocol
\section{Introduction}
Since the publication of Einstein's general relativity in 1914, researchers were captivated by the search for exact solutions of the field equations. Over the past century a myriad of exact solutions were obtained which attempted to explain observations in cosmology and astrophysics. The gravitational field exterior to a static, spherically symmetric star was first obtained by Schwarzschild in 1916. This vacuum solution was followed by the interior Schwarzschild solution which describes the gravitational field of a uniform density sphere [\cite{KS1,KS2}]. Causality is one of the cornerstones of relativity which requires $0 < \frac{dp}{d\rho} < 1$ [\cite{dev1,dev2}]. It is clear that causality is violated at each interior point of the Schwarzschild constant density sphere. This prompted researchers to consider more realistic matter configurations which included inhomogeneous density profiles, anisotropic pressures, electric charge, bulk viscosity and scalar fields. Generalization of the perfect fluid interior matter distribution to include anisotropic stresses has yielded interesting physical characteristics of such models. It was shown that physical properties such as surface tension, compactness and surface red-shift of these stars are  sensitive to the anisotropy parameter [\cite{a0,bowers,gov1,a1}]. The impact of electric charge in compact objects has been widely studied within the context of stability and physical viability. It was shown that the presence of electric charged alters the Buchdahl limit required for stability of a self-gravitating, bounded matter distribution [\cite{buch1,buch2}]. Departure from spherical symmetry has also been pursued in the context of slowly rotating stars and in the description of gravitational waves [\cite{hr1,hr2}]. Various techniques ranging from ad-hoc assumptions, imposition of pressure isotropy, use of an equation of state, use of the condition of conformal flatness, Lie symmetry analysis, to name just a few, were used to solve the field equations [\cite{lie,ivanov}]. While these methods yield solutions, there is no guarantee that the ensuing models are physically viable. An extensive review of exact solutions of the Einstein field equations describing static objects show that a very small subset of these satisfy all the requirements for realistic stellar models [\cite{step}].

A natural question which arises in astrophysics is what happens when a star loses hydrostatic equilibrium and undergoes continued gravitational collapse? Oppenheimer and Snyder tackled this problem by considering a spherically symmetric dust cloud undergoing gravitational collapse [\cite{opp}]. Their model served as a catalyst in understanding end-states of gravitational collapse. The Cosmic Censorship Conjecture which ruled out the formation of naked singularities for collapsing matter configurations with reasonable initial states was shown to be violated under various assumptions [\cite{p1,p2,p3}]. The study of black holes has moved into the observable realm making it a popular research topic [\cite{bh}]. Black hole physics has evolved immensely from the simple Oppenheimer-Snyder dust model to include anisotropic pressures, electromagnetic field, cosmological constant as well as higher dimensions.

In the paper \cite{v} presented an exact solution describing the exterior gravitational field of a radiating star. This solution is a unique solution of the Einstein field equations describing a spherically symmetric atmosphere composed of null radiation. The Vaidya solution made it possible to model dissipative collapse in which the collapsing core radiates energy to the exterior spacetime in the form of a radial heat flux or null radiation. There were several early attempts at modeling a radiating star with a Vaidya exterior. The problem was the matching of the interior and exterior spacetimes across the boundary of the star. The junction conditions required for the smooth matching of a spherically symmetric, shear-free line element to Vaidya's outgoing solution was provided by \cite{san}. It was shown that for a radiating spherical body dissipating energy in the form of a radial heat flux, the pressure on the boundary is proportional to the magnitude of the heat flux. This condition ensures conservation of momentum across the boundary of the collapsing body. Since the publication of the Santos junction conditions, there has been an explosion of models describing dissipative collapse starting with simple solutions and thus rapidly developing into more sophisticated stellar models. The authors \cite{h1,h2,h3,h4,h5} have been instrumental in investigating the nature of collapse with dissipation within a  general framework thus giving researchers rich insights into these problems especially with the inclusion of shear, inhomogeneity and anisotropy. The thermodynamics of radiating stars was developed by Govender and co-workers since the early 1990's. Relaxational effects due to heat dissipation and shear viscosity predict temperature and luminosity profiles which are significantly different from the Eckart theory of thermodynamics [\cite{g1,g2,g3}].
Recently, there has been a resurgence in seeking exact solutions to the Einstein field equations describing static, compact objects by employing the concept of embedding. The Karmarkar condition which needs to be satisfied if the spacetime has to be of class I embedding has been widely used to generate various stellar models describing anisotropic spheres [\cite{karm}]. These models have been shown to satisfy all the stringent stability and physical tests imposed by the behaviour of the thermodynamic and gravitational variables [\cite{k1,k2,k3,jasim20,singh20,sg20,iva20,nayan20}]. Many of these solutions have been reconciled with observational data of compact objects including strange stars, pulsars and neutron stars [\cite{k4,k5,k6,upreti,Ful,pantgd}]. By utilising a quadratic equation of state together with the Karmarkar condition a model for the strange star candidate SAX J1808.4-3658 was obtained. It was shown that this model agrees with observational characteristics of this star. Furthermore, a comparison of the quadratic EoS model with modified Bose-Einstein condensation EoS and linear EoS was carried out \cite{k7}. The Karmarkar condition has also been utilised to model dissipative collapse ensuing from an initially static configuration losing hydrostatic equilibrium and starts to radiate energy to the exterior spacetime. The Karmarkar condition together with the junction condition which represents conservation of momentum across the collapsing boundary determine the temporal and gravitational evolution of the model [\cite{nolene}]. Many of these models indicate their robustness under the scrutiny of physical viability. To this end we employ the Karmarkar condition to seek a model which accurately describes two stellar compact objects, namely, PSR J1614-2230 and SAX J1808.4-3658.

This paper is structured as follows: In section I, we present the Einstein field equations describing the interior spacetime of the stellar model. The Karmarkar and embedding class I conditions are introduced in section III. By adopting a parametric form for one of the metric potentials we generate a stellar model in section IV. The matching of the interior and exterior spacetimes is accomplished in section V. The physical features of the model is discussed in section VI. We investigate the stability of our model in section VII. The paper concludes with a discussion and finding of our main results in section VIII.

\section{The Einstein Field Equations}
The line element within the spherically symmetric anisotropic fluid matter distribution in Schwarzschild coordinates $(x^{i})= (t, r, \theta, \phi)$ is delineated in the following form:
\begin{equation}
ds^2 = e^{\nu(r)}dt^2 - e^{\lambda(r)}dr^2 - r^2(d\theta^2 + \sin^2{\theta}d\phi^2). \label{metric}
\end{equation}
where the gravitational potentials $\nu(r)$ and $\lambda(r)$ are yet unknown.
The energy-momentum tensor for anisotropic matter takes the form
\begin{equation}
T_{jk} = [(p_t+\rho)v_{j}v_k-p_t g_{jk}+(p_r-p_t)\chi_j\chi_k],\label{2}
\end{equation}
where $\rho$, $p_r$ and $p_t$ are the energy density, radial and transverse pressures respectively and $p_t$ is in the perpendicular direction to $p_r$. The normalized 4-velocity vector  $v^{j}=\sqrt {\frac{1}{g_{tt}}}\delta_{t}^{j}$ and the unit spacelike vector $\chi^j= \sqrt {-\frac{1}{g_{rr}}}\delta_{r}^{j}$ along $r$ provide $g_{jk}v^{j}v^k = 1$ and $g_{jk}\chi^j\chi^k=-1$ respectively.

The line element (\ref{metric}) and momentum tensor $T_{jk}$ (\ref{2}) give rise to the following system of equations [\cite{Mnc}]
\begin{eqnarray}\label{g3}
8\pi \rho &=&
\frac{\left(1 - e^{-\lambda(r)}\right)}{r^2} + \frac{\lambda'(r)e^{-\lambda(r)}}{r},\label{g3a} \\ \nonumber \\
8\pi  p_r &=&  \frac{\nu'(r) e^{-\lambda(r)}}{r} - \frac{\left(1 - e^{-\lambda(r)}\right)}{r^2}, \label{g3b}
\end{eqnarray}
\begin{eqnarray}
8\pi  p_t =\frac{e^{-\lambda}}{4}\left(2\nu'' + {\nu'}^2  - \nu'\lambda' + \frac{2\nu'}{r}-\frac{2\lambda'}{r}\right), \label{g3c}
\end{eqnarray}
where $(')$ denotes the derivative with respect to the radial coordinate $r$.

Using the field equations Eqs.(\ref{g3b}) and (\ref{g3c}), the anisotropic factor $(\Delta)$ takes the form
\begin{eqnarray}
\Delta &=& p_t-p_r \nonumber\\
&=& e^{-\lambda}\left[{\nu'' \over 2}-{\lambda'\nu' \over 4}+{\nu'^2 \over 4}-{\nu'+\lambda' \over 2r}+{e^\lambda-1 \over r^2}\right]\label{del}.
\end{eqnarray}
Here we choose the gravitational constant $G$ and speed of sound $c$ to be unity.

\section{The Karmarkar condition}

The Karmarkar condition required for the spacetime to be of class I embedding is
\begin{equation}\label{kar}
R_{1414}=\frac{R_{1212}R_{3434}+R_{1224}R_{1334}}{R_{2323}},
\end{equation}
subject to $R_{2323}\neq 0$ [\cite{R2323}].

The non-zero Riemann tensor components for the line element (\ref{metric}) are
\begin{eqnarray}
R_{1414}&=& -e^{\nu(r)}(\frac{\nu^{''}(r)}{2}+\frac{\nu{'}^2(r)}{4}-\frac{\lambda'(r)\nu'(r)}{4}),\\
R_{2323}&=& -e^{-\lambda(r)}r^2sin^2{\theta}(e^{\lambda(r)}-1),\\
R_{1212}&=& \frac{1}{2}r{\lambda}'(r),\\
R_{3434}&=&-\frac{1}{2}r\sin^2{\theta}\nu'(r)e^{\nu(r)-\lambda(r)}.
\end{eqnarray}

The differential equation derived using the Karmarkar condition (\ref{kar}) assumes the form
\begin{equation}\label{7}
\frac{2\nu^{''}}{\nu'}+\nu'=\frac{\lambda'e^{\lambda(r)}}{e^{\lambda(r)}-1}.
\end{equation}

Solving eqn.(\ref{7}), we find the following relation between $e^{\lambda(r)}$ and $e^{\nu(r)}$
\begin{equation}\label{8}
e^{\lambda(r)}=\Big{(}P+Q\int_{0}^{r}\sqrt{e^{\lambda(r)}-1}dr\Big{)}^2,
\end{equation}
where $P$ and $Q$ are integration constants.

In view of (\ref{del}), the anisotropy of the fluid $\Delta$ [\cite{M0}] is obtained as
\begin{equation}\label{delta}
\Delta=\frac{\nu{'}(r)}{4e^{\lambda(r)}}\Big{[}\frac{2}{r}-\frac{\lambda'(r)}{e^{\lambda(r)}-1}\Big{]}\Big{[}\frac{\nu'(r)e^{\nu(r)}}{2rB^2}-1\Big{]}.
\end{equation}
At this juncture we should point out that when $\Delta = 0$, the only bounded solution simultaneously satisfying the Karmarkar condition and pressure isotropy is the interior Schwarzschild solution. This solution suffers various shortcomings including superluminal speeds within the interior of the fluid. To this end we consider a solution describing an anisotropic fluid distribution which will be taken up in the next section.

\section{A new parametric class solutions}

In this paper, we assumed the following metric potential
\begin{eqnarray} \label{elam}
e^{\lambda(r)}&=&1+a r^2 \alpha_n(r),
\end{eqnarray}
where
\begin{eqnarray*}
\alpha_n(r) &=& \csc ^n\left(b r^2+c\right),
\end{eqnarray*}
and $a$, $b$ and $c$ are positive constants and $n\geq 0$. We have selected $e^{\lambda(r)}$ such that at center $e^{\lambda(r)}=1$, which emphasizes that at the center the tangent 3 space is flat and the Einstein field equations (EFEs) can be integrated. Substituting the $e^{\lambda(r)}$ from (\ref{elam}) in (\ref{8}), we obtain the remaining metric potential $e^{\nu(r)}$ as
\begin{eqnarray} \label{enu}
e^{\nu(r)}&=& \left(P-\frac{Q h_1(r)h_2(r) \sqrt{a\alpha_n(r)}}{4 b}\right)^2,
\end{eqnarray}
where $P$ and $Q$ are integration constants.

Using the metric potentials given by Eqs. (\ref{elam}) and (\ref{enu}), the expressions of $\rho$, $p_r$, $\Delta$ and $p_t$ can be cast as
\begin{eqnarray}
\rho &=& \frac{a \alpha _n(r) \left(r^2 \left(a \alpha _n(r)-2 b n \cot \left(b r^2+c\right)\right)+3\right)}{\left(a r^2 \alpha _n(r)+1\right){}^2} ,\nonumber\\
\end{eqnarray}
\begin{eqnarray}
p_r&=& \frac{h_2(r) \sqrt{a\alpha_n(r)}}{h_3(r) \left(a r^2 \alpha_n(r)+1\right)},
\end{eqnarray}

\begin{eqnarray}
\Delta&=& \frac{h_5(r) r^2 \left(2 b h_6(r)-h_7(r)\right)}{h_8(r) \left(1+a r^2 \alpha_n(r)\right)^2},\nonumber \\
\end{eqnarray}
\begin{eqnarray}
p_t=p_r+\Delta,
\end{eqnarray}
where
\begin{eqnarray*}
h_1(r) &=& \, _2F_1\left(\frac{1}{2},\frac{n+2}{4};\frac{3}{2};\cos ^2\left(b r^2+c\right)\right),\\
h_2(r) &=& \sin \left(2 \left(b r^2+c\right)\right) \sin ^2\left(b r^2+c\right)^{\frac{n-2}{4}},\\
h_3(r) &=& 2 P b \sqrt{a \alpha _n}-a Q h_1(r) \sqrt{\alpha _n} \cos \left(b r^2+c\right)\\&&-4 b Q,\\
h_4(r)&=& \sqrt{a} Q h_1(r)\cos \left(b r^2+c\right)-2 P b \\
h_5(r) &=& a \alpha _n(r)+b n \cot \left(b r^2+c\right)\\
h_6(r)&=& a P \alpha _n(r)-Q \sqrt{a \alpha _n(r)}\\
h_7(r)&=&aB h_4(r) \cos \left(b r^2+c\right) \csc ^{\frac{n}{2}}\left(b r^2+c\right)\\
h_8(r)&=& 2 P b-\sqrt{a} Q h_1(r) \cos \left(b r^2+c\right)
\end{eqnarray*}

The mass function $m(r)$, gravitational red-shift $z(r)$ and compactification factor $u(r)$ at the surface and within the interior of the stellar system are given by
\begin{eqnarray}
m(r)&=& \frac{a r^3 \alpha _n(r)}{2 \left(a r^2 \alpha _n(r)+1\right)}, \\
z(r) & = & \frac{1}{P-\frac{Q h_1(r) h_2(r) \sqrt{a \alpha _n(r)}}{4 b}}-1,\\
u(r)&=&\frac{m(r)}{r}=\frac{a r^2 \alpha _n(r)}{2 \left(a r^2 \alpha _n(r)+1\right)}.
\end{eqnarray}


\section{Matching of interior and exterior spacetime over the boundary}
To determine the constants $a, ~b,~ c,~P,~ Q$ appearing in our class of solutions, the interior metric must be matched smoothly across the boundary with the exterior Schwarz-schild solution
\begin{eqnarray}
ds^2 &=& \Big(1-{2M\over r}\Big) dt^2-\Big(1-{2M\over r}\Big)^{-1}dr^2\nonumber\\&-& r^2(d\theta^2+\sin^2 \theta d\phi^2). \label{ext}
\end{eqnarray}
By comparing the interior solution (\ref{metric}) with exterior solution (\ref{ext}) at the boundary $r=R$ (Darmois-Isreali conditions), we obtain
\begin{eqnarray}
e^{\nu_b} &=& 1-{2M \over R}\nonumber \\ &=&  \left(P+\frac{Q \left(n \sqrt{1-\gamma}+2 b R^2+2 c\right) \sqrt{a\alpha_n(R)}}{b \left(n^2+4\right)}\right)^2,\nonumber\\ \label{bou1}
\end{eqnarray}

\begin{eqnarray}
e^{-\lambda(r)_b} = 1-{2M \over R}
= \frac{1}{1+a R^2 \alpha_n(R)}, \label{bou2}
\end{eqnarray}
\begin{eqnarray}
p_r(R) = 0 \label{bou3}.
\end{eqnarray}
With the help of the boundary conditions (\ref{bou1}-\ref{bou3}),  we obtain
\begin{eqnarray}
a &=&  -\frac{2 M \csc ^{-n}\left(b R^2+c\right)}{R^2 (2 M-R)},\label{a}\\
P &=&\frac{\sqrt{1-\frac{2 M}{R}} \left(a h_1(R) \cos \left(b R^2+c\right) \sqrt{\alpha_{n}(R)}+4 b\right)}{4 b},\nonumber\\ \label{P}\\ 
Q &=&\frac{1}{2} \sqrt{1-\frac{2 M}{R}} \sqrt{a \csc ^n\left(b R^2+c\right)}, \label{Q}
\end{eqnarray}
where $\gamma = \left(b R^2+c\right)^2$.

The constants $b$ and $c$ are free parameters and are selected such a way that all the physical properties of the assumed stars for a suitable range of $n$ are well-behaved and satisfy the Darmois-Israel conditions. The values of $P$ and $Q$ are expressed in Eqs. (\ref{P}) and (\ref{Q}) respectively.

\section{Discussion of physical features for well-behaved solutions}

\subsection{Geometrical regularity}

The metric potentials (geometrical parameters) for the stars PSR J1614-2230 and SAX J1808.4-3658 for the range of $n$ mentioned in Table 1 at the center $(r=0)$, give the values $e^\nu|_{r=0}=$ positive constant and $e^{-\lambda(r)}|_{r=0}=1$. This shows that the metric potentials are regular and free from geometric singularities inside the stars. Also, both metric potentials $e^\nu(r)$ and  $e^{-\lambda(r)}$ are monotonically increasing and decreasing respectively, with  $r$ (Fig.\ref{metric}).\\

\subsection{Viable trends of physical parameters}
\subsubsection{Density and pressure trends}
\begin{figure}
\centering
\includegraphics[height=2in,width = 2.8in]{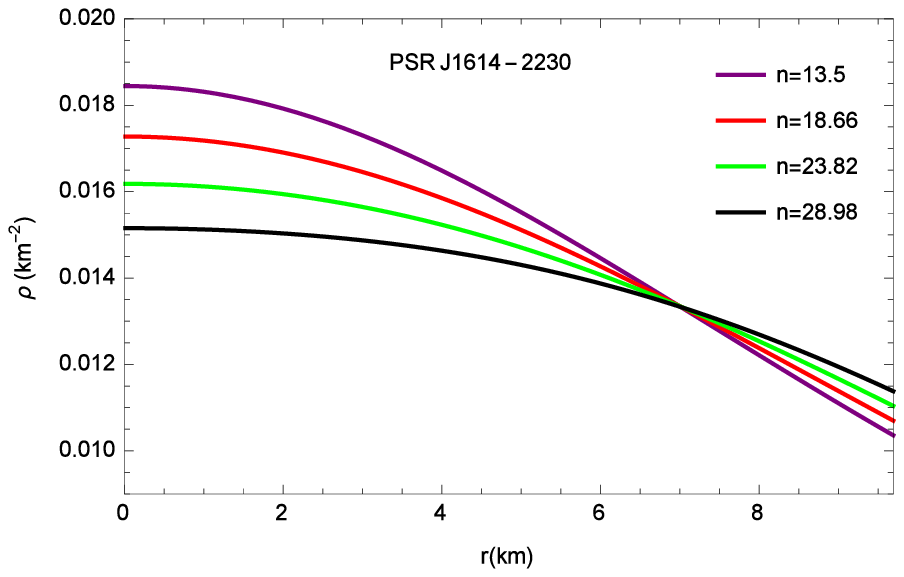}\includegraphics[height=2in,width = 2.8in]{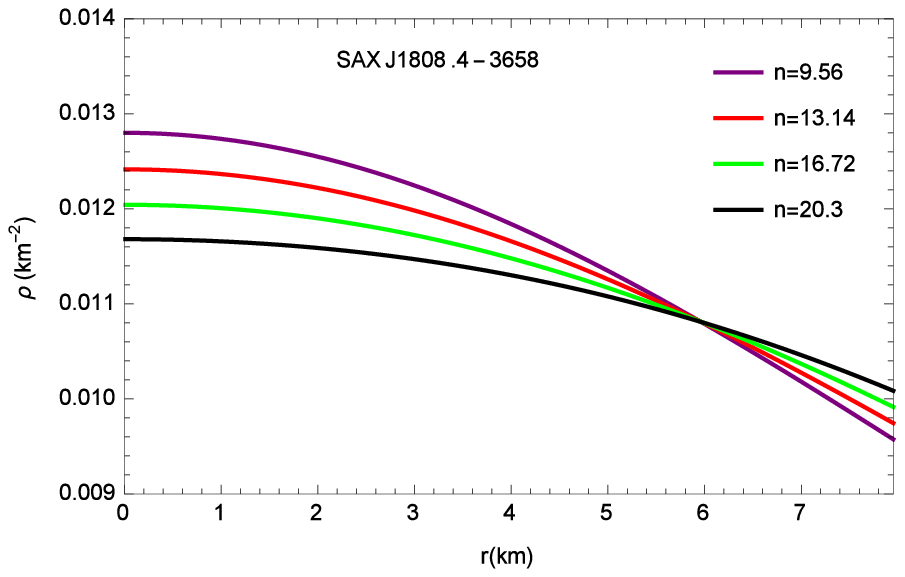}
\caption{Variation of $\rho$ with $r$ for (i)PSR J1614-2230 with mass $M=1.97 M_{\odot}$ and radius $R=9.69$km for the models $n=13.5, ~18.66, ~ 23.82, ~28.98$; (ii) SAX J1808.4-3658 with mass $M=0.9 M_{\odot}$ and radius $R=7.951$km for the models $n=~9.56,~13.14,~16.72,~20.3$ and the values of $b = 0.0001/km^2$, $c=2.5$.}
\label{rho}
\end{figure}

\begin{figure}
\centering
\includegraphics[height=2in,width = 2.8in]{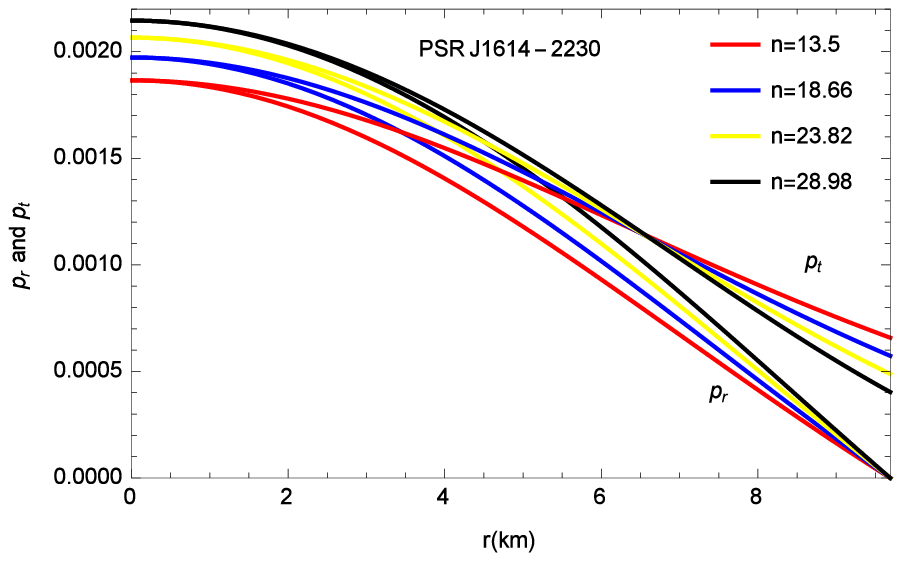}\includegraphics[height=2in,width = 2.8in]{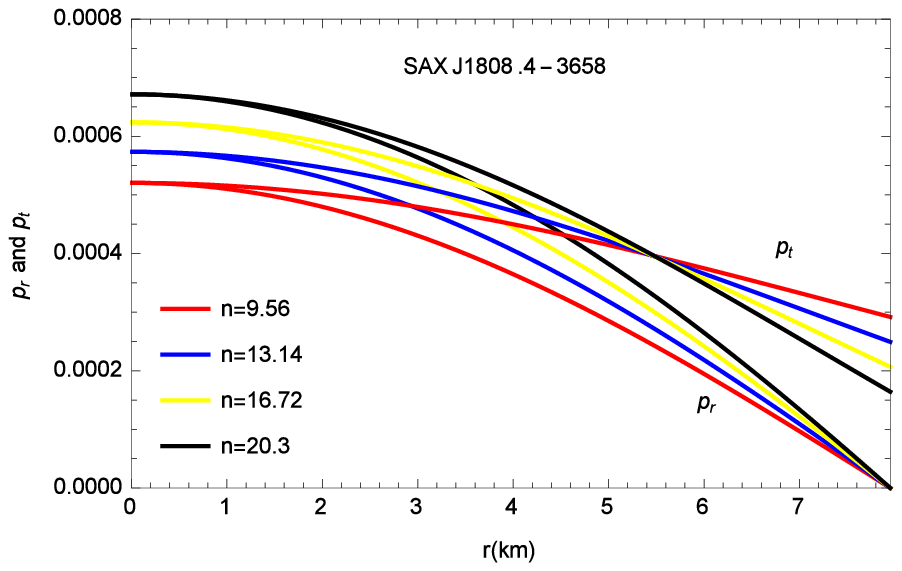}
\caption{Variation of $p_r$, $p_t$ with $r$ for (i)PSR J1614-2230 with mass $M=1.97 M_{\odot}$ and radius $R=9.69$km for the models $n=13.5, ~18.66, ~ 23.82, ~28.98$; (ii) SAX J1808.4-3658 with mass $M=0.9 M_{\odot}$ and radius $R=7.951$km for the models $n=~9.56,~13.14,~16.72,~20.3$ and the values of $b = 0.0001/km^2$, $c=2.5$.}
\label{prpt}
\end{figure}

The matter density $\rho$, radial pressure $p_r$ and transverse pressure $p_t$ for stars PSR J1614-2230 and SAX J1808.4-3658 are non-negative inside the stars and monotonically decrease from center to surface of these stars for the range of $n$ mentioned in Table 1(Fig.\ref{rho},Fig.\ref{prpt}) [\cite{ZN,I1}]. 

\subsubsection{Relation between pressure-density ratios (Equation of state)}
\begin{figure}
\centering
\includegraphics[height=2in,width = 2.8in]{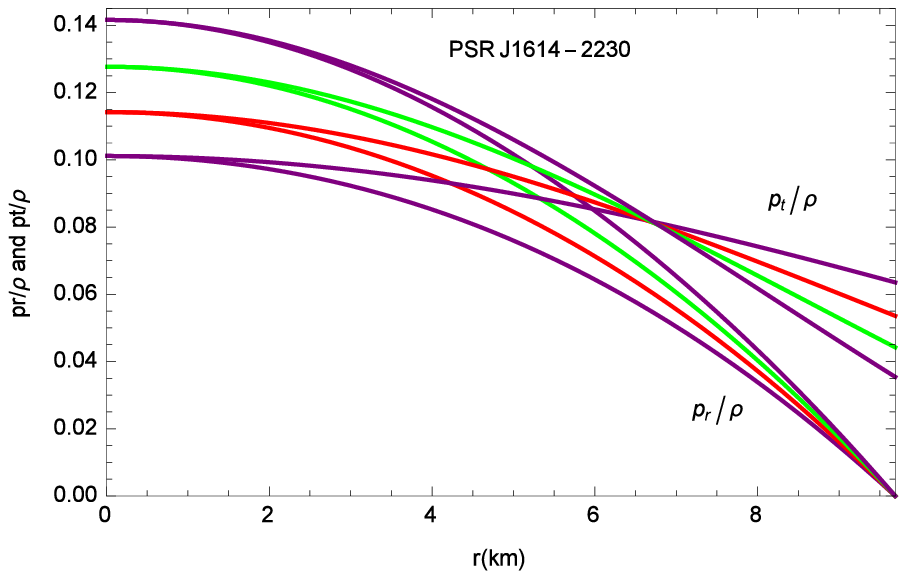}\includegraphics[height=2in,width = 2.8in]{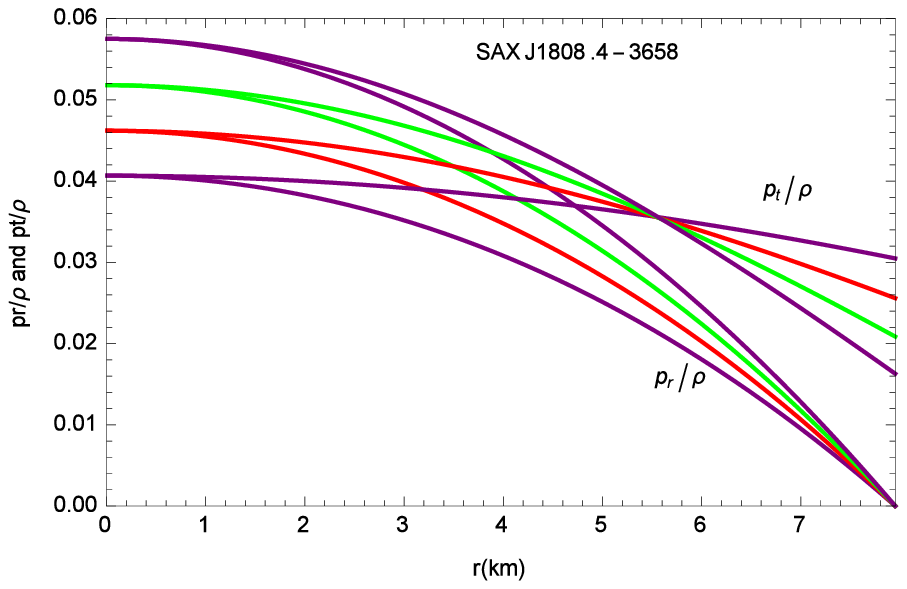}
\caption{Variation of $p_r/\rho$ and $p_t/\rho$ with $r$ for (i)PSR J1614-2230 with mass $M=1.97 M_{\odot}$ and radius $R=9.69$km for the models $n=13.5, ~18.66, ~ 23.82, ~28.98$; (ii) SAX J1808.4-3658 with mass $M=0.9 M_{\odot}$ and radius $R=7.951$km for the models $n=~9.56,~13.14,~16.72,~20.3$ and the values of $b = 0.0001/km^2$, $c=2.5$.}
\label{prptrho}
\end{figure}

\begin{figure}
\centering
\includegraphics[height=2in,width = 2.8in]{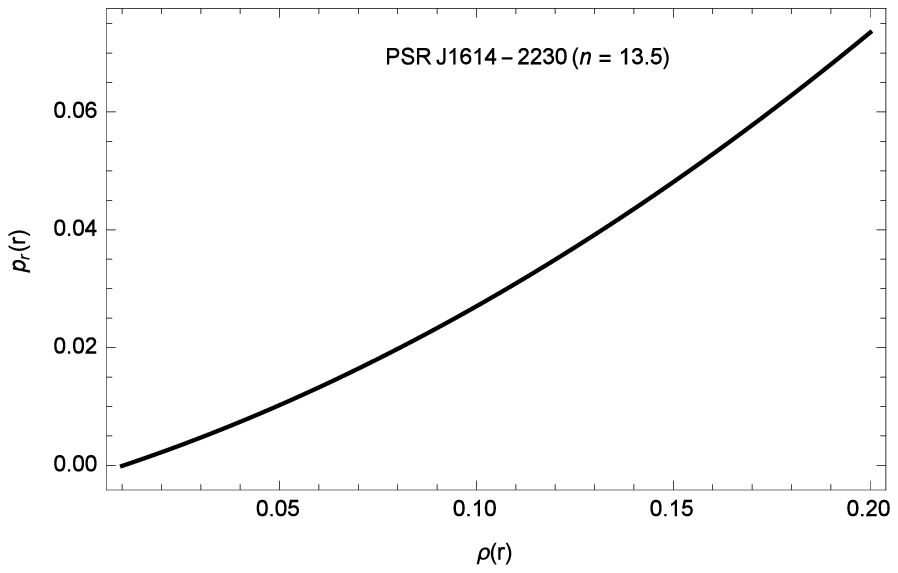}
\includegraphics[height=2in,width = 2.8in]{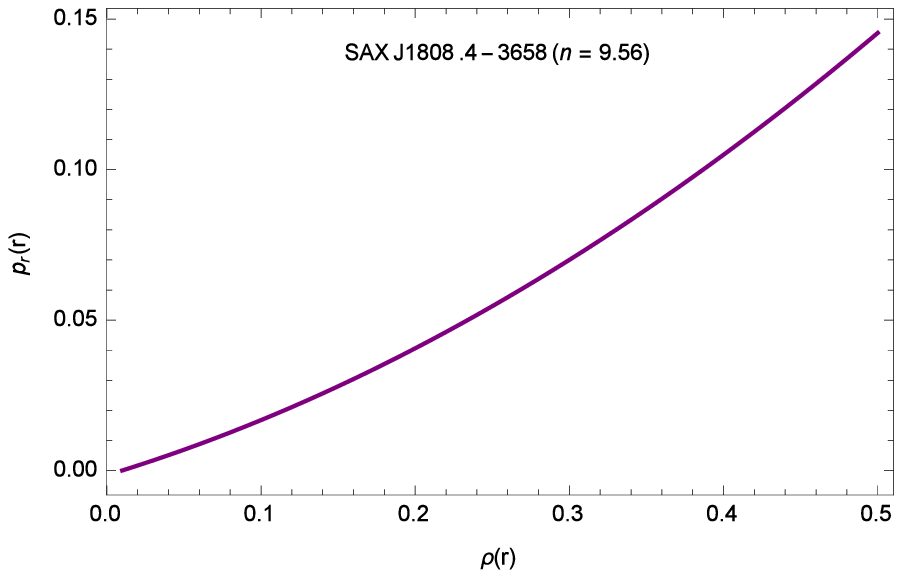}
\caption{Variation of equation of state parameters with $\rho$ for (i)PSR J1614-2230 with mass $M=1.97 M_{\odot}$ and radius $R=9.69$km for the model $n=13.5$; (ii) SAX J1808.4-3658 with mass $M=0.9 M_{\odot}$ and radius $R=7.951$km for the model $n=~9.56$ and the values of $b = 0.0001/km^2$, $c=2.5$.}
\label{eos}
\end{figure}

We plot the graphs of the equation of state parameters ($p_r/\rho, ~p_t/\rho$) to establish some connection between density and the pressures. Using the trend of plots, we establish a linear, quadratic or CFL EoS for our  model. An example of starting off with the metric functions and then establishing an EoS is the classic paper by \cite{Mukherjee}. In this work they show that the Vaidya-Tikekar geometry leads to a linear EoS. 
From the plots of figures, we observe the decreasing trend of pressure to density ratios with $r$ Fig.(\ref{prptrho}) for both the stars PSR J1614-2230 and SAX J1808.4-3658 for the range of $n$ mentioned in Table 1. Based on the trends of the plots, we calculate equation of state (EOS) for neutron star PSR J1614-2230 as 
\begin{eqnarray}
p_r &=& 0.861538 \rho^2 +0.206369 \rho -0.00223306,\\
p_r &=& 69.1848\rho^2 -1.27803 \rho +0.00560289, 
\end{eqnarray} for $n=13.5$, $n=28.98$ respectively and for the strange star SAX J1808.4-3658 as 
\begin{eqnarray}
p_r &=& 0.276979 \rho^2 +0.155325\rho -0.00151322,\\
p_r &=& 48.6746 \rho^2 -0.639035\rho + 0.00149093, 
\end{eqnarray} 
for $n=9.56$, $n=20.3$ respectively, using the method of least of square technique (elaborated in appendix). The profiles of equation of state for PSR J1614-2230 ($n=13.5$ ), SAX J1808.4-3658 ($n=9.56$)  are exhibited in the Fig.(\ref{eos}). The trends of EOS for other values of $n$ in their corresponding ranges of the stars remain same as in the Fig.(\ref{eos}).

\subsubsection{Mass-radius relation, red-shift and compactification factor}
\begin{figure}
\centering
\includegraphics[height=2in,width = 2.8in]{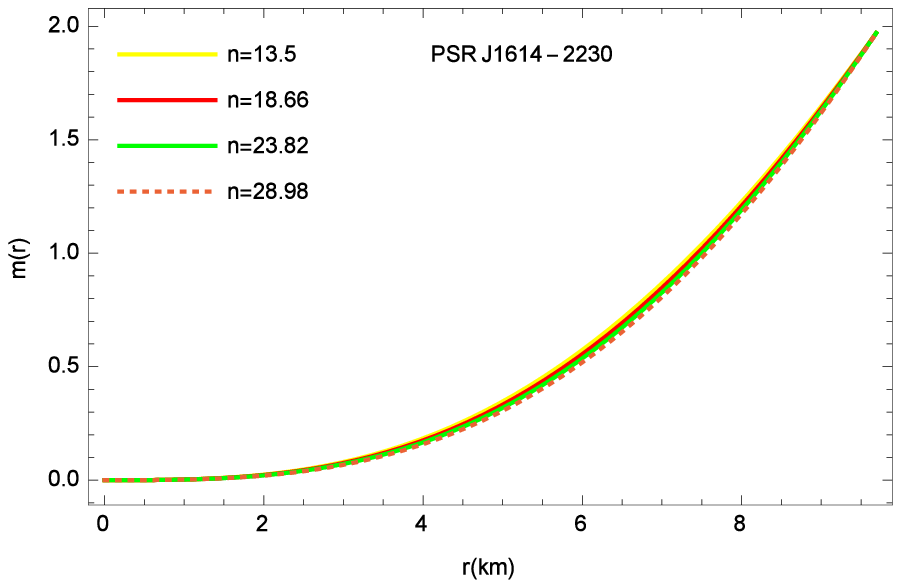}
\includegraphics[height=2in,width = 2.8in]{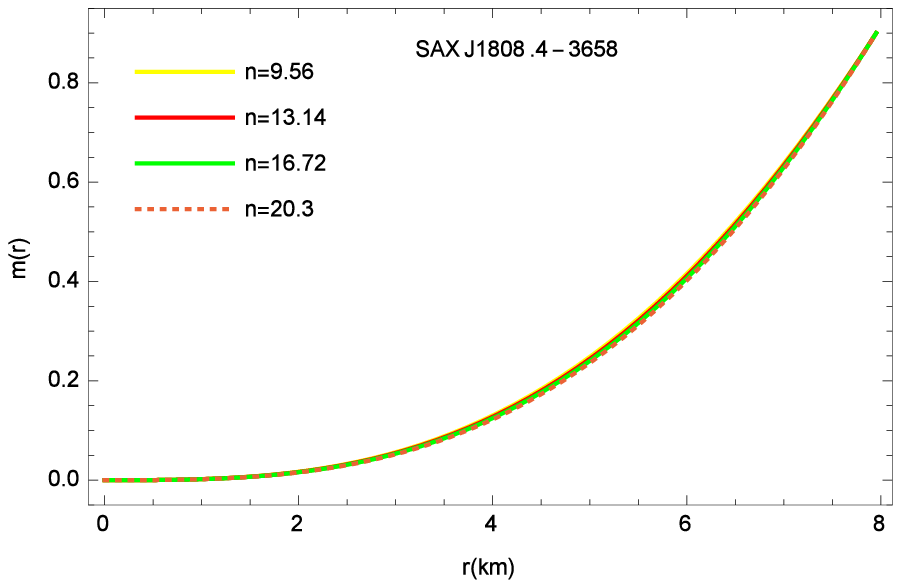}
\caption{Variation of mass ($m(r)$) with $r$ for (i)PSR J1614-2230 with mass $M=1.97 M_{\odot}$ and radius $R=9.69$km for the models $n=13.5, ~18.66, ~ 23.82, ~28.98$; (ii) SAX J1808.4-3658 with mass $M=0.9 M_{\odot}$ and radius $R=7.951$km for the models $n=~9.56,~13.14,~16.72,~20.3$ and the values of $b = 0.0001/km^2$, $c=2.5$.}
\label{mass}
\end{figure}

\begin{figure}
\centering
\includegraphics[height=2in,width = 2.8in]{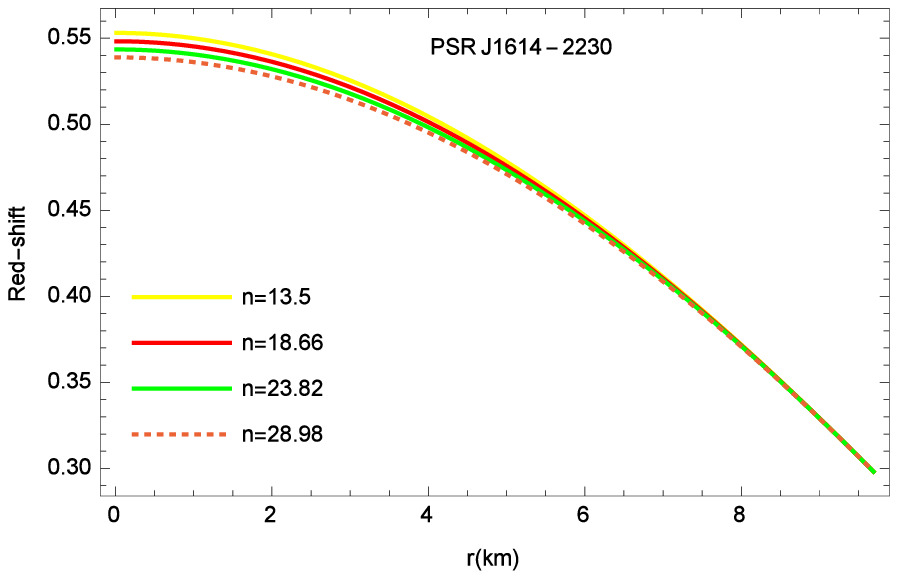}
\includegraphics[height=2in,width = 2.8in]{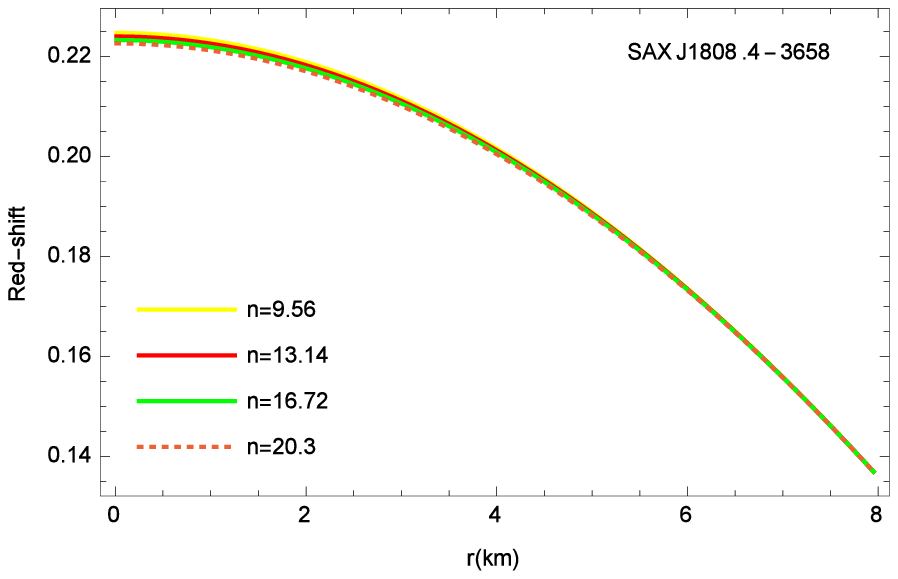}
\caption{Variation of red-shift with $r$ for (i)PSR J1614-2230 with mass $M=1.97 M_{\odot}$ and radius $R=9.69$km for the models $n=13.5, ~18.66, ~ 23.82, ~28.98$; (ii) SAX J1808.4-3658 with mass $M=0.9 M_{\odot}$ and radius $R=7.951$km for the models $n=~9.56,~13.14,~16.72,~20.3$ and the values of $b = 0.0001/km^2$, $c=2.5$.}
\label{zr}
\end{figure}

\begin{figure}
\centering
\includegraphics[height=2in,width = 2.8in]{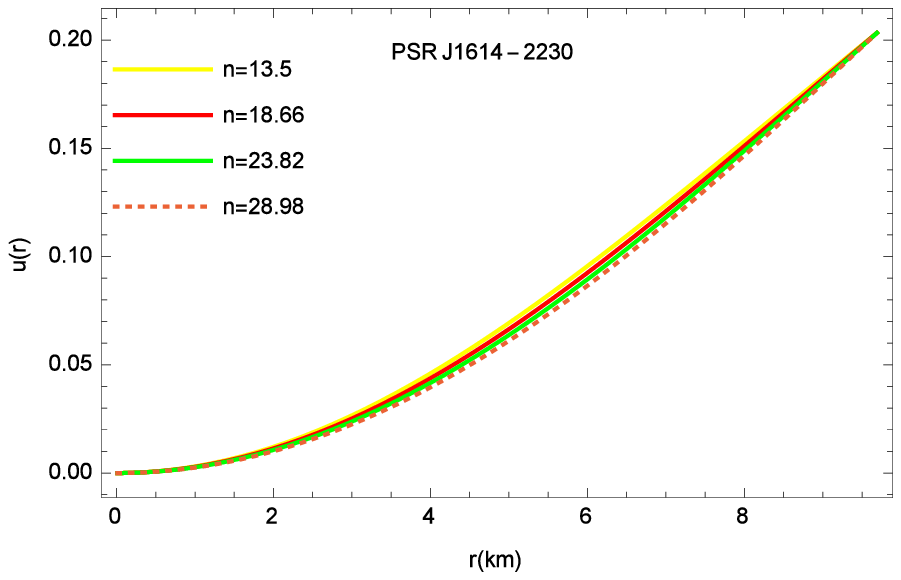}
\includegraphics[height=2in,width = 2.8in]{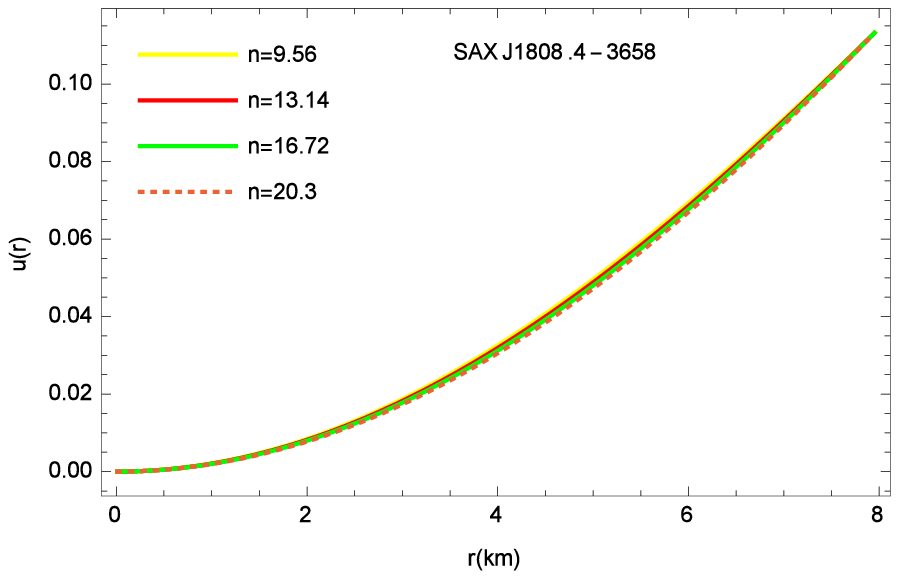}
\caption{Variation of the compactification factor $u(r)$ with $r$ for (i)PSR J1614-2230 with mass $M=1.97 M_{\odot}$ and radius $R=9.69$km for the models $n=13.5, ~18.66, ~ 23.82, ~28.98$; (ii) SAX J1808.4-3658 with mass $M=0.9 M_{\odot}$ and radius $R=7.951$km for the models $n=~9.56,~13.14,~16.72,~20.3$ and the values of $b = 0.0001/km^2$, $c=2.5$.}
\label{u}
\end{figure}

The mass function $m(r)$ and gravitational  red-shift $z(r)$ function of stars PSR J1614-2230 and SAX J1808.4-3658 for the range of $n$ mentioned in Table 1 are increasing  and decreasing respectively with $r$. The variation of $m(r)$ and $z(r)$ is shown in Figs.(\ref{mass},\ref{zr}). Also, compactification parameter $u(r)$ for both the stars are increasing functions with $r$, shown in Fig.(\ref{u}) and lies within the Buchdahl limit [\cite{Bu}].

\subsubsection{Anisotropic parameter}
In Fig.(\ref{delta}), the radial pressures ($p_r$) coincides with tangential pressure ($p_t$) at the center of stars PSR J1614-2230 and SAX J1808.4-3658 for the range $n$ mentioned in Table 1, i.e, pressure anisotropies vanish at the center, $\Delta(0) = 0$ and increase outwards [\cite{bowers,I1}].

\section{Physical Stability analysis}
\subsection{Zeldovich's condition}

The values of $p_r$, $p_t$ and $\rho$ at the center are given by
\begin{eqnarray}\label{p0}
8\pi p_{rc}  = 8\pi p_{tc}~~~~~~~~~~~~~~~~~~~~~~~~~~~~~~~~~~~~~~~~~~~~~~~~ \nonumber \\ =  a \csc ^n(c)\frac{\left(-2 P b \sqrt{a \csc ^n(c)}+4 b Q+\beta _1 \beta _2 Q\right)}{ \left(2 P b \sqrt{a \csc ^n(c)}-\beta _1 \beta _2 Q\right)}>0, \nonumber \\
\end{eqnarray}
and
\begin{eqnarray}
8\pi \rho_c &=& 3 a \csc ^n(c) >0 ~~\textit{if}~~a > 0 . \label{r0}
\end{eqnarray}

Using Zeldovich's condition [\cite{ZN}], i.e., $ {p_{rc}}/ {\rho_c} \leq 1$, we get

\begin{eqnarray}\label{p0r0}
\frac{-2 P b \sqrt{a \csc ^n(c)}+4 b Q+\beta _1 \beta _2 Q}{3 \left(2 P b \sqrt{a \csc ^n(c)}-\beta _1 \beta _2 Q\right)}\le 1,
\end{eqnarray}

In view of (\ref{r0}) and (\ref{p0r0}), we get the following inequality
\begin{eqnarray}
\frac{2Ab\sqrt{a\csc ^n(c)}}{4b+\beta_1\beta_2}\leq  \frac{Q}{P} \leq \frac{2Ab\sqrt{a\csc ^n(c)}}{b+\beta_1\beta_2},
\end{eqnarray}
where
\begin{eqnarray*}
\beta_1&=& \, _2F_1\left(\frac{1}{2},\frac{n+2}{4};\frac{3}{2};\cos ^2(c)\right),\\
\beta_2&=& a \cos (c) \sin ^{\frac{n}{2}}(c) \csc ^n(c).
\end{eqnarray*}

\subsubsection{Hererra cracking stability of an anisotropic fluid sphere}

The Hererra cracking method [\cite{cracking}] is used to analyze the stability of anisotropic stars under radial perturbations. We also employ the concept of cracking due to \cite{AHN} to analyze potentially stable regions within the stellar configuration by subjecting our model to the condition $-1< v^2_t-v^2_r \leq 0$
\begin{eqnarray}
\frac{dp_t}{d\rho}=\frac{dp_r}{d\rho}+\frac{d\Delta}{d\rho}=\frac{dp_r}{d\rho}+\frac{d\Delta}{d\rho}\frac{dr}{d\rho},
\end{eqnarray}
\begin{eqnarray}
v^2_r-v^2_t=-\frac{d\Delta}{d\rho}\frac{dr}{d\rho}.
\end{eqnarray}

\begin{figure}
\centering
\includegraphics[height=2in,width = 2.8in]{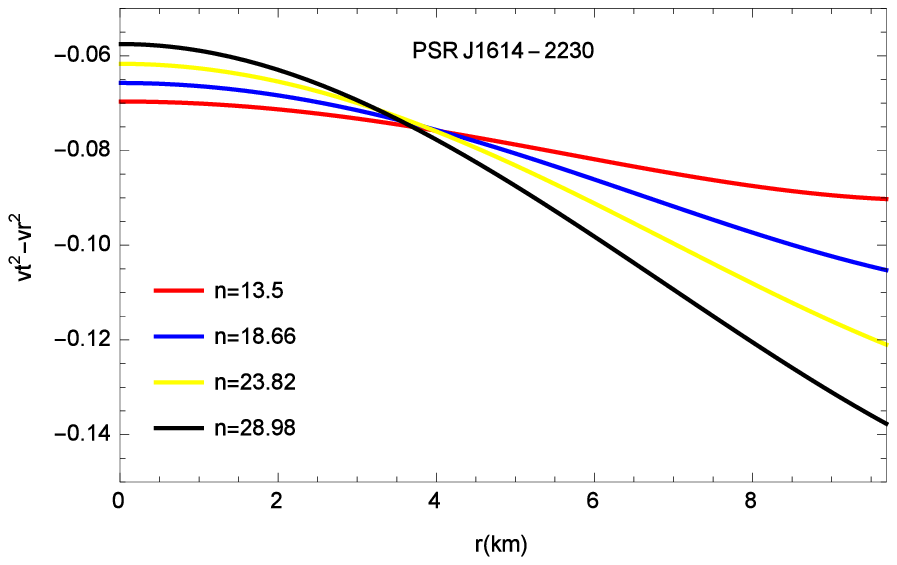}
\includegraphics[height=2in,width = 2.8in]{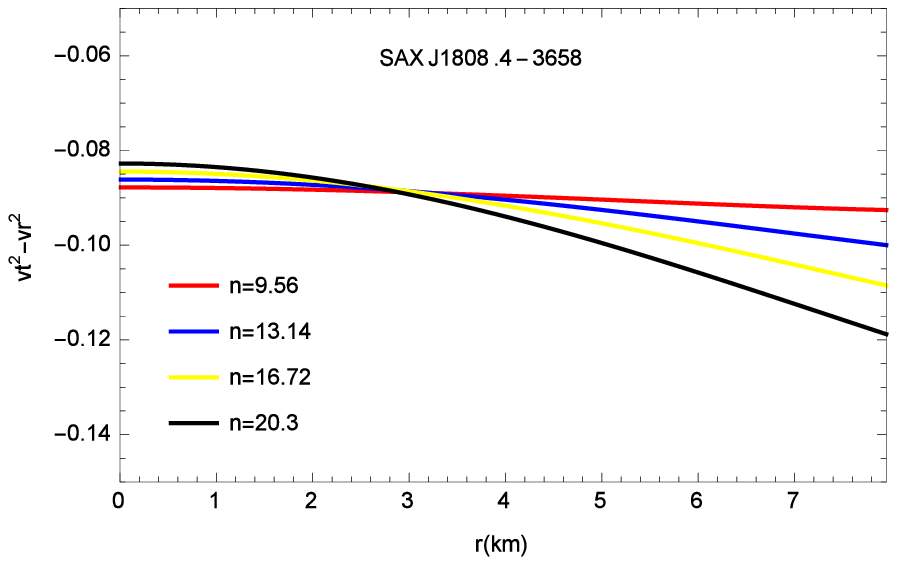}
\caption{Variation of ${v_t}^2-{v_r}^2$ with $r$ for (i)PSR J1614-2230 with mass $M=1.97 M_{\odot}$ and radius $R=9.69$km for the models $n=13.5, ~18.66, ~ 23.82, ~28.98$; (ii) SAX J1808.4-3658 with mass $M=0.9 M_{\odot}$ and radius $R=7.951$km for the models $n=~9.56,~13.14,~16.72,~20.3$ and the values of $b = 0.0001/km^2$, $c=2.5$.}
\label{vt2-vr2}
\end{figure}

For a physically feasible model of anisotropic fluid sphere the radial and transverse velocities of sound should be less than 1, which are referred to as causality conditions in the literature.
The profiles of $v^2_r$ and $v^2_t$ of stars PSR J1614-2230 and SAX J1808.4-3658 for the range $n$ mentioned in Table 1 are given in Fig.(\ref{vr2vt2}),  which shows that $0< v^2_r \leq 1$ and $0< v^2_t \leq 1$ everywhere within the stellar configuration. Therefore, both the speeds satisfy the causality conditions and monotonically decreasing nature. Here, we use the Herrera cracking method [\cite{cracking}] for analyzing the stability of anisotropic stars under the radial perturbations. Using the concept of cracking, \cite{AHN} gave the idea that the region of the anisotropic fluid sphere where $-1< v^2_t-v^2_r \leq 0$ is potentially stable. Fig.(\ref{vt2-vr2}) clearly depicts that our model is potentially stable inside the both stars PSR J1614-2230 and SAX J1808.4-3658 for the range $n$ mentioned in Table 1.

\subsubsection{Bondi stability condition for adiabatic index}
For a relativistic anisotropic sphere the stability depends
on the adiabatic index $\Gamma_r$, the ratio of two specific heats,
defined by  \cite{HH},
\begin{center}
$\Gamma_r = \frac{\rho + p_r} {p_r} \frac{\partial p_r}{\partial\rho}$.
\end{center}

\cite{Bo} suggested that for a stable Newtonian sphere, $\Gamma$ value should be greater than $\frac{4}{3}$.  For an anisotropic relativistic sphere the stability condition is given by \cite{Ch},

\begin{center}
$\Gamma > \frac{4}{3}+ \big[\frac{4 (p_{t0}-p_{r0})}{3 |p^{'}_{r0}|r}+\frac{\rho_{0}p_{r0}}{2 |p^{'}_{r0}|}r\big]$,\\
\end{center}
where $p_{r0}, p_{t0}$ and $\rho_0$ represent the initial radial pressure, tangential
pressure and energy density respectively in static equilibrium. The first and last term inside the square brackets represent the anisotropic and relativistic corrections respectively. Moreover, both the quantities are positive and  increase the unstable range of $\Gamma$.

\cite{Chandra1} established a condition on $\Gamma$ to  study the stability of interior of Schwarzschild metric and it is defined as
\begin{eqnarray}
\Gamma &>& \Gamma_{cr}= \frac{4}{3}+ \frac{19}{42}{(2\delta)},\label{gcr}
\end{eqnarray}
where $\delta$ is compactification factor and $\Gamma_{cr}$ is the critical adiabatic index which is determined from neutral configuration. 

\cite{mousta} suggested that in the interior of fluid sphere $\Gamma_{cr}$ should linearly depend on the pressure and density rations at center and $\Gamma > \Gamma_{cr}$ . For stable Newtonian sphere, Bondi and Chandrasekhar suggested that $\Gamma >\frac{4}{3}$ [\cite{Bo,Chandra1,Chandra2}].

The present class of models satisfy Bondi, Chandrasekhar, Moustakidis conditions  for both the compact stars PSR J1614-2230 and SAX J1808.4-3658 for the range of $n$ mentioned in Table 1 and $\Gamma_{cr}$ linearly depend on the ratio $\frac{p_r(0)}{\rho(0)}$ .

\subsubsection{Energy conditions}
\begin{figure}
\centering
\includegraphics[height=2in,width = 2.8in]{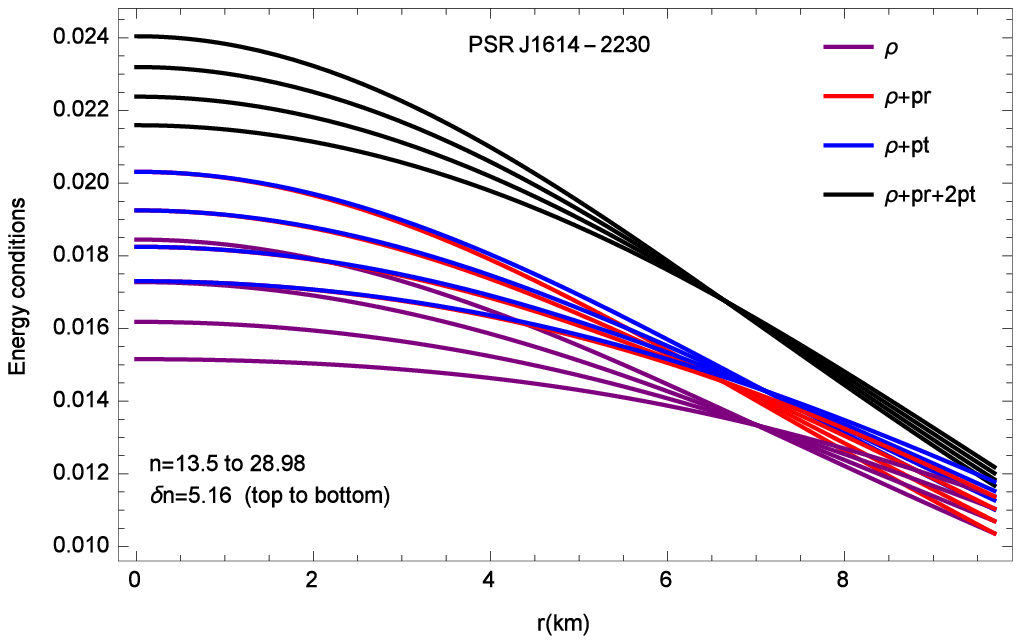}\includegraphics[height=2in,width = 2.8in]{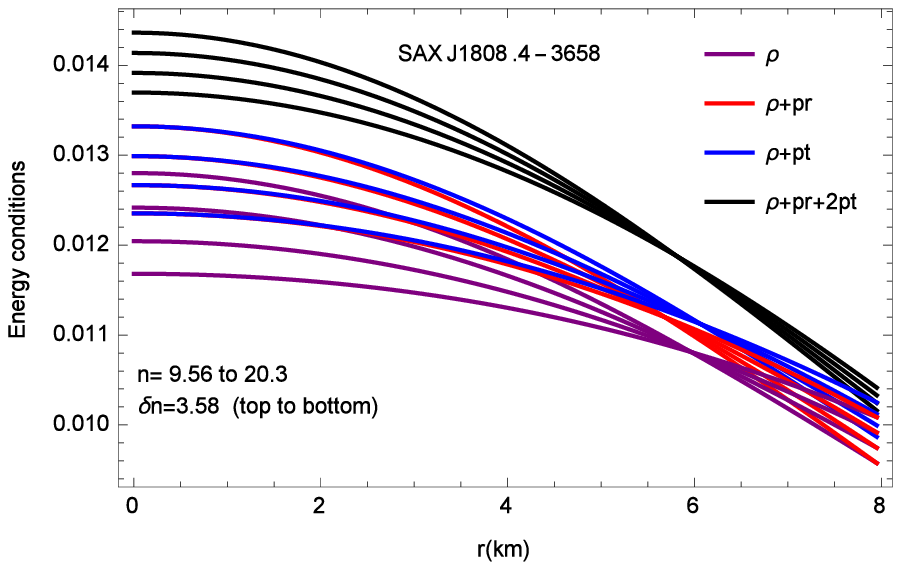}
\caption{Variation of energy conditions with  $r$ for (i)PSR J1614-2230 with mass $M=1.97 M_{\odot}$ and radius $R=9.69$km for the models $n=13.5, ~18.66, ~ 23.82, ~28.98$; (ii) SAX J1808.4-3658 with mass $M=0.9 M_{\odot}$ and radius $R=7.951$km for the models $n=~9.56,~13.14,~16.72,~20.3$ and the values of $b = 0.0001/km^2$, $c=2.5$.}
\label{enercond}
\end{figure}

For a physically stable static model the interior of the star should satisfy (i) null energy condition $\rho + p_r \geq 0$ (NEC) (ii) weak energy conditions $\rho + p_r \geq 0$, $\rho \geq 0$ (WEC$_r$) and $\rho + p_t \geq 0$, $\rho \geq 0$ (WEC$_t$) and (iii) strong energy condition $\rho+p_r+2p_t \geq 0$ (SEC) [\cite{Mn}]. The profiles of energy conditions i.e. NEC, WEC, SEC are displayed in Fig.(\ref{enercond}) and our models satisfy all the energy conditions for both the stars PSR J1614-2230 and SAX J1808.4-3658 for the range $n$ mentioned in Table 1.

\subsection{Tolman-Oppenheimer-Volkoff condition for equilibrium under three forces}
The Tolman-Oppenheimer-Volkoff (TOV) equation [\cite{TOV}] for anisotropic fluid matter distribution is given as
\begin{eqnarray}
-{M_g(r)(\rho+p_r) \over r^2}~e^{(\lambda(r)-\nu(r))/2}-{dp_r \over dr}+{2\Delta (r) \over r}=0, \label{tove}
\end{eqnarray}
where $F_g$, $F_h$, $F_a$ are  gravitational, hydrostatic and  anisotropic forces respectively and $M_g(r)$ is the gravitational mass can be obtained from the Tolman-Whittaker formula
\begin{eqnarray}
M_g(r) &=& {1\over 2}~r^2 \nu'(r) e^{(\nu(r)-\lambda(r))/2}.
\end{eqnarray}

The TOV equation (\ref{tove}) can be expressed in the following  balanced force equation
\begin{equation}
F_g+F_h+F_a=0, \label{forc}.
\end{equation}

\begin{figure}
\centering
\includegraphics[height=2in,width = 2.8in]{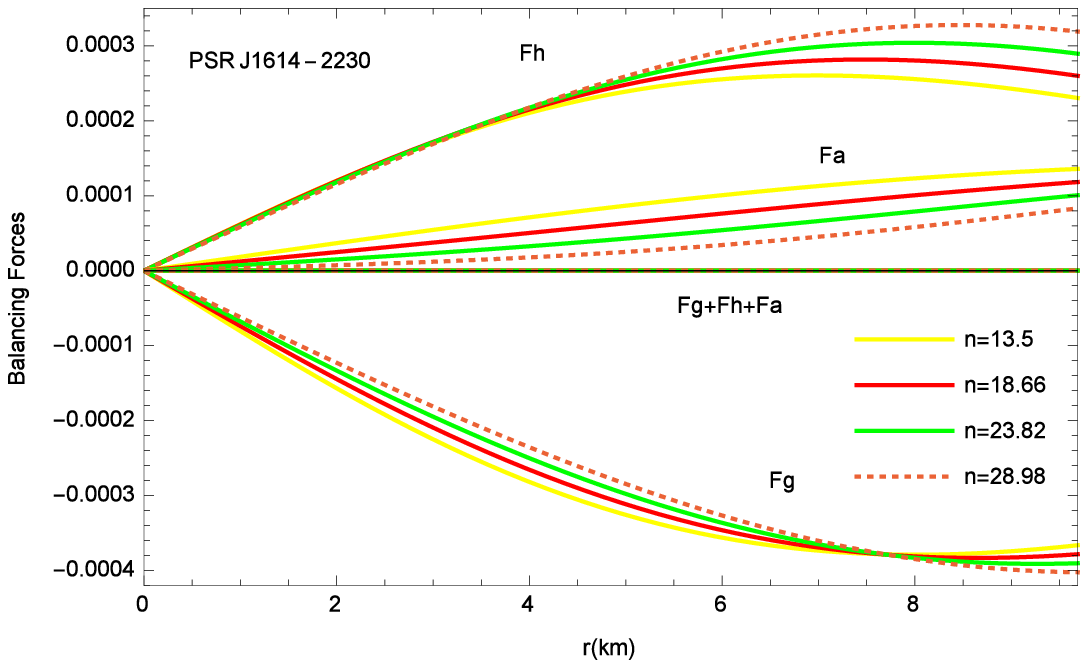}
\includegraphics[height=2in,width = 2.8in]{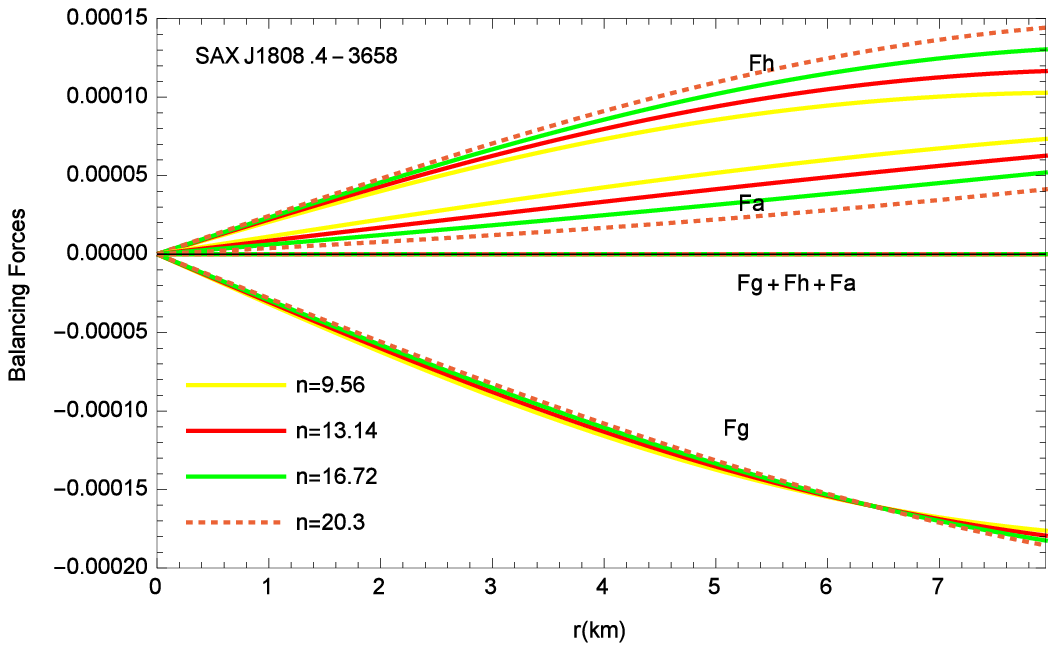}
\caption{Variation of balancing forces $F_a, ~F_g,~ F_a,~ F_a+F_g+F_h$ with  $r$ for (i)PSR J1614-2230 with mass $M=1.97 M_{\odot}$ and radius $R=9.69$km for the models $n=13.5, ~18.66, ~ 23.82, ~28.98$; (ii) SAX J1808.4-3658 with mass $M=0.9 M_{\odot}$ and radius $R=7.951$km for the models $n=~9.56,~13.14,~16.72,~20.3$ and the values of $b = 0.0001/km^2$, $c=2.5$.}
\label{balfor}
\end{figure}

In an equilibrium state the three forces $F_g$, $F_h$ and $F_a$  satisfy TOV equation. The profiles of the three forces of the stars PSR J1614-2230, SAX J1808.4-3658 are exhibited in Fig.(\ref{balfor}) and in which $F_g$ overshadows the other two forces $F_h$ and $F_a$ such that the system to be in a static equilibrium.

\subsection{Harrison-Zeldovich-Novikov Static stability criterion}
\begin{figure}
\centering
\includegraphics[height=2in,width = 2.8in]{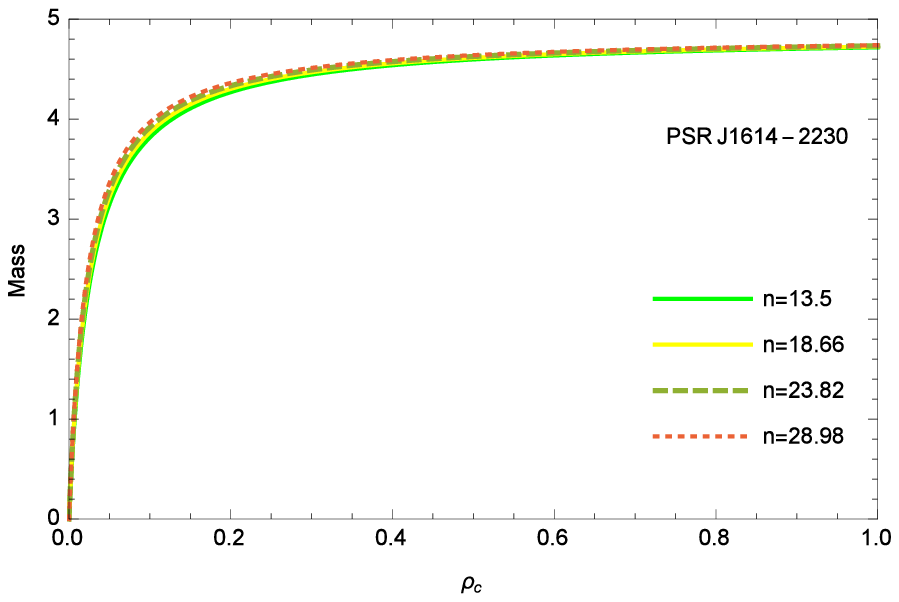}
\includegraphics[height=2in,width = 2.8in]{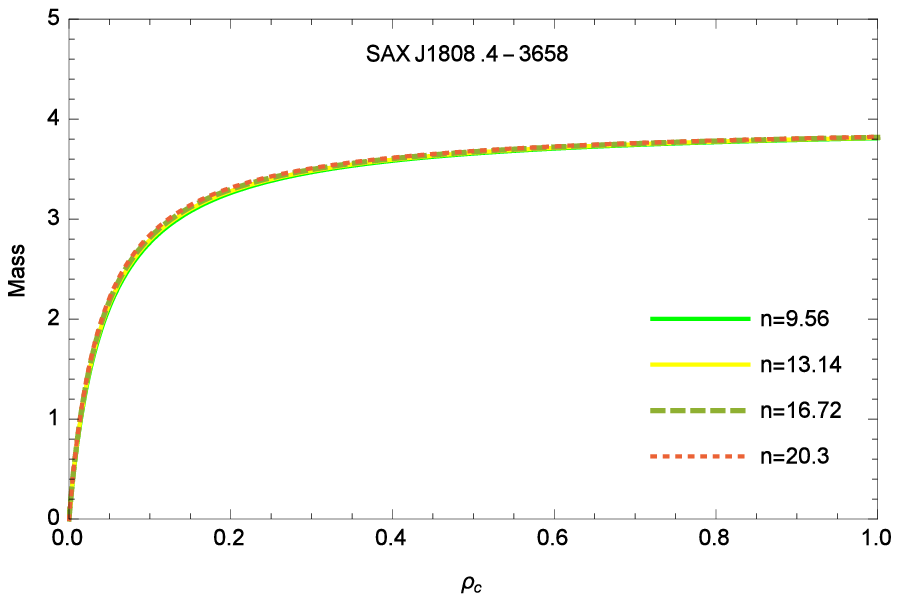}
\caption{Variation of mass with central density $\rho_c$ for (i)PSR J1614-2230 with mass $M=1.97 M_{\odot}$ and radius $R=9.69$km for the models $n=13.5, ~18.66, ~ 23.82, ~28.98$; (ii) SAX J1808.4-3658 with mass $M=0.9 M_{\odot}$ and radius $R=7.951$km for the models $n=~9.56,~13.14,~16.72,~20.3$ and the values of $b = 0.0001/km^2$, $c=2.5$.}
\label{mrhoc}
\end{figure}

The Harrison-Zeldovich-Novikov static stability criteria  for non-rotating spherically symmetric equilibrium stellar models  provides that the mass of compact stars must be an increasing function of its central density under small radial pulsation i.e.
\begin{eqnarray} \label{stc}
\frac{\partial M}{\partial\rho_c}>0.
\end{eqnarray}
This criteria ensures that the model is static and stable. It was proposed by \cite{Har} and \cite{ZN} independently for stable stellar models. With the help of  (\ref{r0}) and total mass
\begin{eqnarray}
M=m(R)=\frac{a R^3 \csc ^n\left(b R^2+c\right)}{2 \left(a R^2 \csc ^n\left(b R^2+c\right)+1\right)},
\end{eqnarray}
The expression of the mass in terms of the central density is given by
\begin{eqnarray*}
M(\rho_c)= \frac{\rho  R^3 \csc ^{-n}(c) \csc ^n\left(b R^2+c\right)}{2 \left(\rho  R^2 \csc ^{-n}(c) \csc ^n\left(b R^2+c\right)+3\right)}.
\end{eqnarray*}
Also,
\begin{eqnarray*}
\frac{\partial M}{\partial\rho_c}= \frac{R^3 \csc ^{-n}(c) \csc ^n\left(b R^2+c\right)}{6 \left(\frac{1}{3} \rho  R^2 \csc ^{-n}(c) \csc ^n\left(b R^2+c\right)+1\right)^2}>0,
\end{eqnarray*}
satisfies (Fig.\ref{mrhoc}) the static stability criterion (\ref{stc}).

The Harrison-Zeldovich-Novikov condition is satisfied for both the stars PSR J1614-2230 and SAX J1808.4-3658 for the range $n$ mentioned in Table 1.

\section{Discussion and Conclusion}
\begin{table*}[t]
\caption{The variation in physical parameters, i.e., central adiabatic index, central density, central red-shift, surface red-shift and compactness factor for different models of (i)PSR J1614-2230 with mass $M=1.97 M_{\odot}$ and radius $R=9.69$km for parameters $n=13.5, ~18.66, ~ 23.82, ~28.98$; (ii) SAX J1808.4-3658 with mass $M=0.9 M_{\odot}$ and radius $R=7.951$km for parameters $n=~9.56,~13.14,~16.72,~20.3$ for the values of $b = 0.0001/km^2$, $c=2.5$, $G=6.67 \times 10^{-11} m^3 kg^{-1} s^{-2}$, $M_{\odot}= 2\times 10^{30} kg$ and $C=3\times 10^8 m s^{-1}$.}
{\begin{tabular}{|p{3cm}|p{1.4cm}|p{1.4cm}|p{1.4cm}|p{1.2cm}|p{1.2cm}|p{1.2cm}|p{1.2cm}|p{1.2cm}|}
\\\hline
 & $n=13.5$ & $n=18.66$& $n=23.82$ & $n=28.98$& $n=9.56$& $n=13.14$& $n=16.72$& $n=20.3$
 \\\hline
 Central adiabatic index($\Gamma_c$) &2.5881  &3.2254  & 4.5  & 8.0523 & 4.2634 & 5.1352  & 6.7296  & 9.76  \\ \hline
 Central density($\rho_c$ $g/cm^3\times 10^{14}$) &  4.8075& 4.3597 & 3.9575 & 3.5959&  3.6093& 3.4397  & 3.2792  & 3.127\\\hline
 Central radial Pressure ($P_{r_c}$) ($dyne/cm^2 $ $\times 10^{34}$) & 9.008 & 9.5265 & 9.9761 & 10.362 & 2.5136  &2.7694  & 3.0125 & 3.2432\\ \hline
Central red-shift ($z_c$)& 0.5531& 0.5482 & 0.5435 & 0.5389 & 0.22474& 0.22402  & 0.22332  & 0.22262 \\ \hline
Surface red-shift ($z_b$) & 0.29815 & 0.29815 & 0.29815 & 0.29815 &  0.13694&  0.13694 & 0.13694 &0.13694\\ \hline
Compactness factor $\frac{GM}{C^2 R}$
& 0.30134 & 0.30134 & 0.30134 & 0.30134 & 0.16777 & 0.16777  & 0.16777 & 0.16777\\ \hline
\end{tabular}}
\end{table*}
Our aim in this paper is to use the Karmarkar condition (which is purely geometric) to establish a physically viable stellar model (albeit a toy model). Toy models are important as they give a sense of the behaviour of the various physical and thermodynamical properties of the star and assist in setting up numerical codes and simulations.

In this paper, we have explored a new parametric class of solutions for anisotropic matter distribution to model the compact star PSR J1614-2230 and strange star SAX J1808.4-3658 by invoking the Karmarkar condition and adopting a form for one of the metric potentials, $e^{\lambda(r)}$. We find a range for one of the parameters, $n$ for the both stars such that the solutions are well behaved for particular choices of the free constants $b,~ c$.  We have analyzed all the geometrical and physical properties of these two stars and verified the physically viability of the solutions for the same range of $n$. 

The graphs of the two stars for different models i.e. (i) $n=13.5, ~18.66, ~ 23.82, ~28.98$ for PSR J1614-2230; (ii) $n=~9.56,~13.14,~16.72,~20.3$ for SAX J1808.4-3658 for parameters values of $b = 0.0001/km^2$, $c=2.5/km^2$ are plotted to find the range of $n$ such that the solutions are well behaved. Furthermore, we concluded that the range of well behaved solutions for PSR J1614-2230 is $n= 13.5~ \mbox{to}~ 28.98$ and for SAX J1808.4-3658 is $n=9.56 ~\mbox{to} ~20.3$ corresponding to same parameter values $b, ~c$.

\begin{figure}
\centering
\includegraphics[height=2in,width = 2.8in]{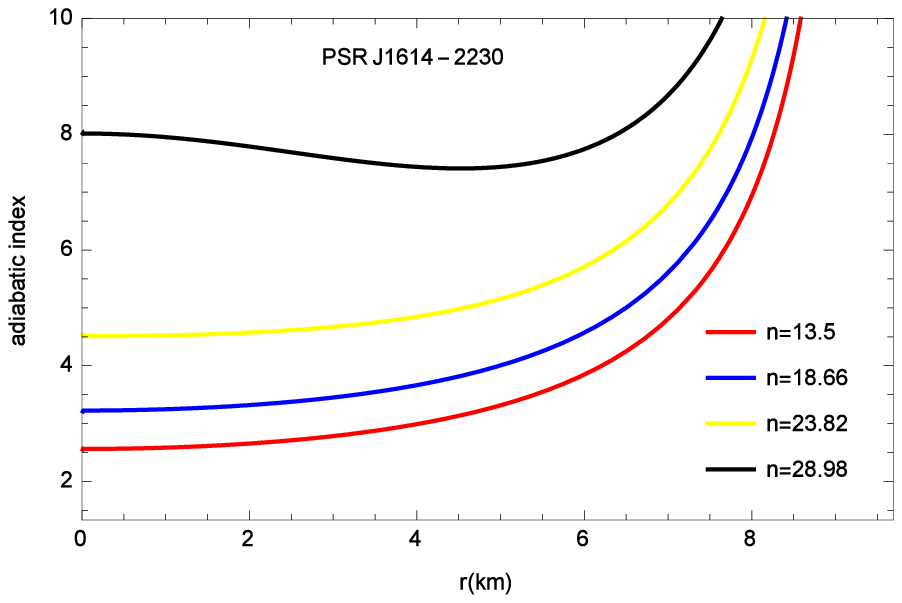}
\includegraphics[height=2in,width = 2.8in]{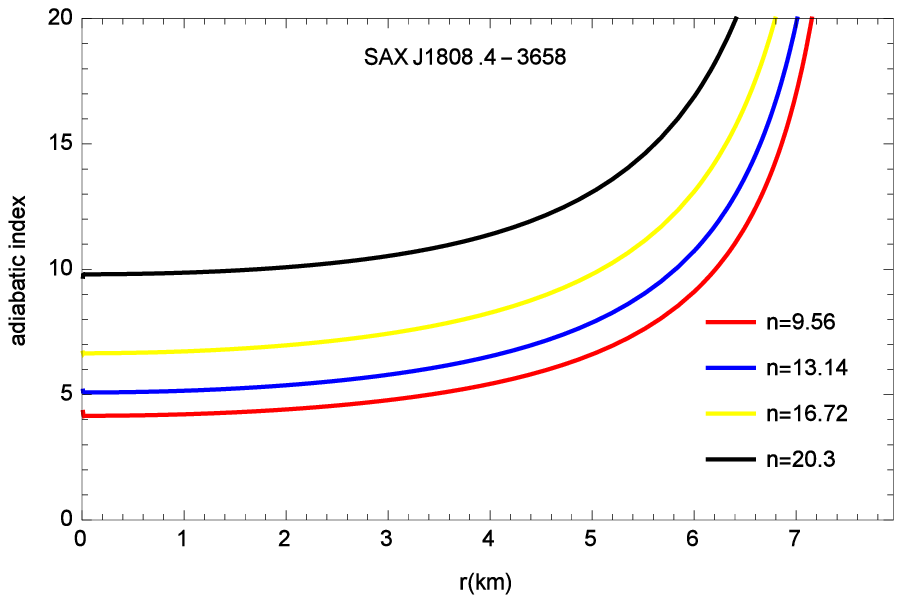}
\caption{Variation of $\Gamma(r)$ with $r$ for (i)PSR J1614-2230 with mass $M=1.97 M_{\odot}$ and radius $R=9.69$km for the models $n=13.5, ~18.66, ~ 23.82, ~28.98$; (ii) SAX J1808.4-3658 with mass $M=0.9 M_{\odot}$ and radius $R=7.951$km for the models $n=~9.56,~13.14,~16.72,~20.3$ and the values of $b = 0.0001/km^2$, $c=2.5$.}
\label{gamma}
\end{figure}

For any value in the range of $n$ the geometrical parameters $(e^{-\lambda(r)}$ and $e^{\nu(r)}$) are decreasing and increasing respectively throughout interior of the stars and both curves meet at their boundary (Fig.\ref{metric}). The physical parameters such as density, radial and tangential pressures, pressures to density ratios, red-shift, both the velocities in that range of $n$ are non-negative  at the center  and  monotonically  decreasing  from center to surface of the stars Figs. (\ref{rho},\ref{prpt},\ref{prptrho},\ref{zr},\ref{vr2vt2}). Physical parameters mass, compactification factor, anisotropy and adiabatic index  are increasing outward which is required for a physically viable stellar configuration Figs.(\ref{mass},\ref{u},\ref{delta},\ref{gamma}).

Our models satisfy all the stability conditions for the two stars for any value of $n$ in that range, i.e, Herrera cracking condition ($-1<v_t^2-v_r^2<0$, $0<v_r^2,~ v_t^2<1$), Bondi condition ($\Gamma>4/3$), Zeldovich’s condition ($0<\frac{p_r}{\rho},~\frac{p_r}{\rho}<1$) and Harrison-Zeldovich-Novikov criterion ($\frac{\partial M}{\partial\rho_c}>0$) Figs.(\ref{vt2-vr2},\ref{gamma},\ref{mrhoc}). For the same range of $n$ of the both stars the present models hold all the energy conditions ($\rho>0$, $\rho+p_r>0$, $\rho+p_t>0$, $\rho+p_r+2p_t>0$) which are required for a physically viable configuration (Fig.\ref{enercond}). Furthermore, our models represent a static anisotropic stellar fluid in equilibrium configuration as the gravitational force, the hydrostatic force  and  the  anisotropic  force  are  acting in the interior stars through  the TOV equation are counter-balancing each other (Fig.\ref{balfor}).

The physical quantities i.e., central adiabatic index ($\Gamma_c$), central density ($\rho_c$), central pressure ($p_{r_c}$), central red-shift ($z_c(r)$), surface red-shift ($z_s(c)$) and compactness factor ($u(r)=\frac{GM}{cR^2}$) are given in Table 1. From Table 1 we conclude that the larger the value of $n$, the central adiabatic index and  central pressure are increasing, whereas the central density and central red-shift are decreasing with increasing the value of $n$. Other physical parameters i.e. compactification factor and red-shift at the surface remain constant for any value of the range $n$. This work has provided a family of parametric solutions of the Einstein field equations obeying the Karmarkar condition. We show that these solutions are sufficiently useful to model compact objects and predict their observed stellar characteristics within very good approximation. 

\section*{Appendix: Generating function}
All the spherically symmetric solutions can be generated from the two generating functions given by \cite{HOP}. The two primitive generating functions $\eta(r)$ and $\Pi(r)$ are given as
\begin{eqnarray}
e^{\nu(r)}=&e^{\big{[}\int{(2\eta{(r)}-\frac{2}{r}})dr\big{]}},\quad \Pi(r)=8\pi(p_r-p_t).
\end{eqnarray}
The two generating functions pertaining to the present class of solutions are obtained as
\begin{eqnarray*}
\eta(r) = ~~~~~~~~~~~~~~~~~~~~~~~~~~~~~~~~~~~~~~~~~~~~~~~~~~~~~~~~~~~~~~~~~~~~\\  \frac{\sqrt{a} Q h_1(r) \cos \left(b r^2+c\right)-2 b \left(\sqrt{a} Q r^2 \csc ^{\frac{n}{2}}\left(b r^2+c\right)+P\right)}{r \left(\sqrt{a} Q h_1(r) \cos \left(b r^2+c\right)-2 P b\right)}
\end{eqnarray*}
and
\begin{eqnarray*}
\Pi(r)&=&8\pi(p_r-p_t)=-8\pi\Delta.
\end{eqnarray*}

\section*{Appendix: Equation of state}
\begin{figure}
\centering
\includegraphics[height=2in,width = 2.8in]{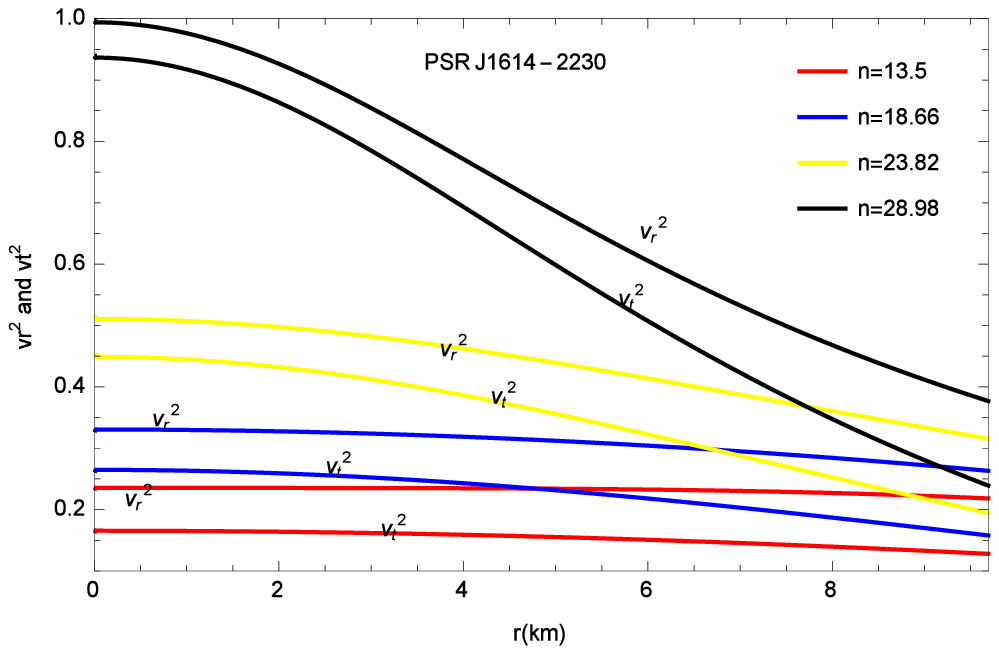}
\includegraphics[height=2in,width = 2.8in]{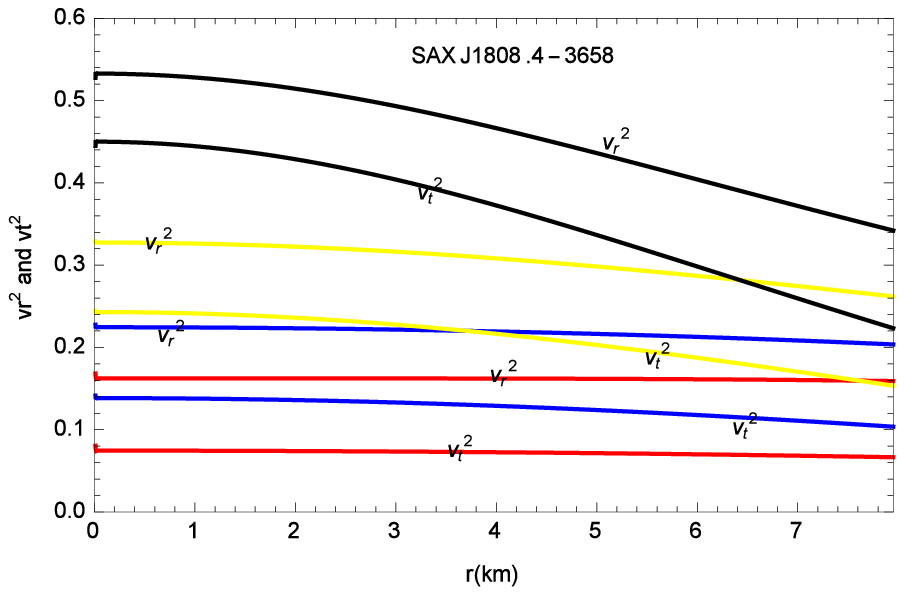}
\caption{Variation of $v_r^2$, $v_t^2$ with $r$ for (i)PSR J1614-2230 with mass $M=1.97 M_{\odot}$ and radius $R=9.69$km for the models $n=13.5, ~18.66, ~ 23.82, ~28.98$; (ii) SAX J1808.4-3658 with mass $M=0.9 M_{\odot}$ and radius $R=7.951$km for the models $n=~9.56,~13.14,~16.72,~20.3$ and the values of $b = 0.0001/km^2$, $c=2.5$.}
\label{vr2vt2}
\end{figure}

The equation of state is defined as the relation between radial pressure ($p_r$) and density ($\rho$) within the star. Since the presence of cumbersome transformation of $p_r$ in terms of $\rho$, here we use curve fitting technique of approximation to get equation of state. Further, from Fig.(\ref{vr2vt2}), we  observe that the plot of $v_r=\sqrt{\frac{dp_r}{d\rho}}$ is not a straight line (i.e. $\frac{dp_r}{d\rho}$ is not a constant), therefore, it is necessary that the relation between $p_r$ and $\rho$ is parabolic in nature. Consequently, in order to get the equation of state we consider the curve fitting method for quadratic form
\begin{eqnarray}
 p_r(r) &=&  U +T \rho(r) +S  \rho^2(r),\label{eos1}\\
\Sigma p_r(r) &=& 11 U+ T \Sigma \rho(r) + S \Sigma \rho^2(r),\label{eos2}\\
\Sigma \rho(r)~ p_r(r) &=& U \Sigma \rho(r) +T \Sigma \rho^2(r) + S \Sigma \rho^3(r) \label{eos3}
\\
\Sigma \rho^2(r) ~p_r(r) &=&  U \Sigma \rho^2(r) +T  \Sigma \rho^3(r) +S \Sigma \rho^4(r),\label{eos4}
\end{eqnarray}
where, $r$ varies from central to boundary of the star. To find the curve via least square method, we consider the points with the differences $0.969$, $0.7951$ for PSR J1614-2230, SAX J1808.4-3658 respectively.
Solving the Eqns.(\ref{eos2},\ref{eos3},\ref{eos4}) for $S$, $T$, $U$ and substituting the values  in Eq.({\ref{eos1}}) , we get required equation of state.

\onecolumn

\begin{figure}
\centering
\includegraphics[height=2in,width = 2.8in]{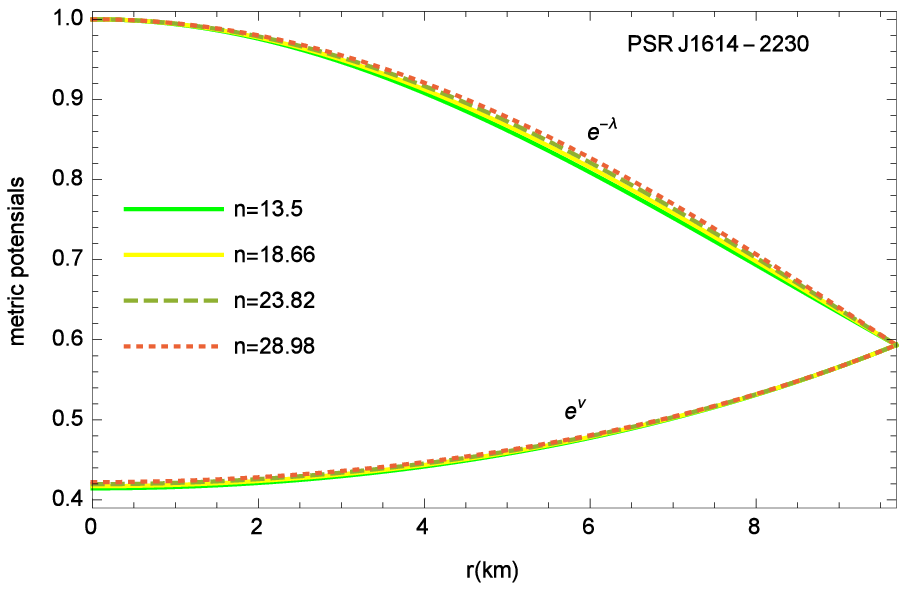}
\includegraphics[height=2in,width = 2.8in]{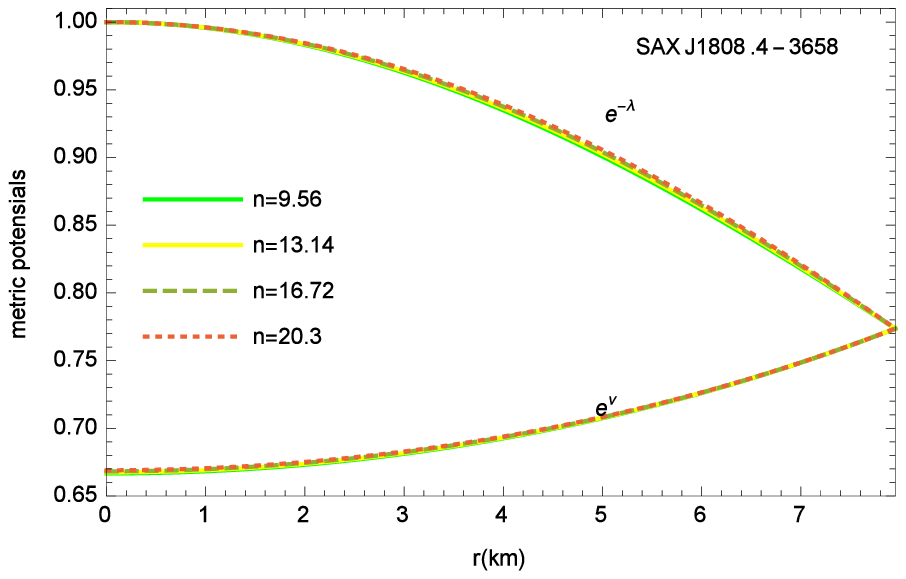}
\caption{Variation of $e^{-\lambda(r)}$, $e^{\nu(r)}$ with $r$ for (i) PSR J1614-2230 with mass $M=1.97 M_{\odot}$ and radius $R=9.69$km for the models $n=13.5, ~18.66, ~ 23.82, ~28.98$; (ii) SAX J1808.4-3658 with mass $M=0.9 M_{\odot}$ and radius $R=7.951$km for the models $n=~9.56,~13.14,~16.72,~20.3$ and the values of $b = 0.0001/km^2$, $c=2.5$.}
\label{metric}
\end{figure}

\begin{figure}
\centering
\includegraphics[height=2in,width = 2.8in]{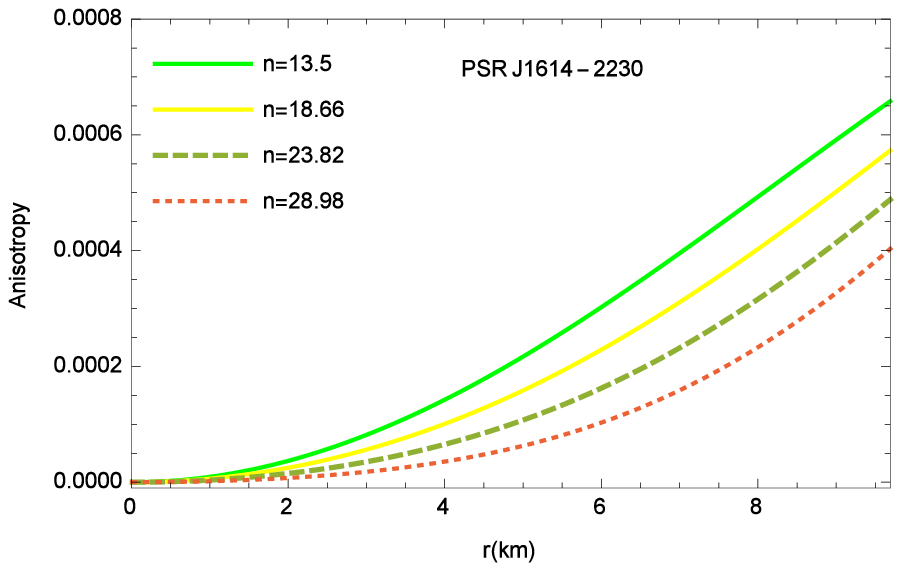}
\includegraphics[height=2in,width = 2.8in]{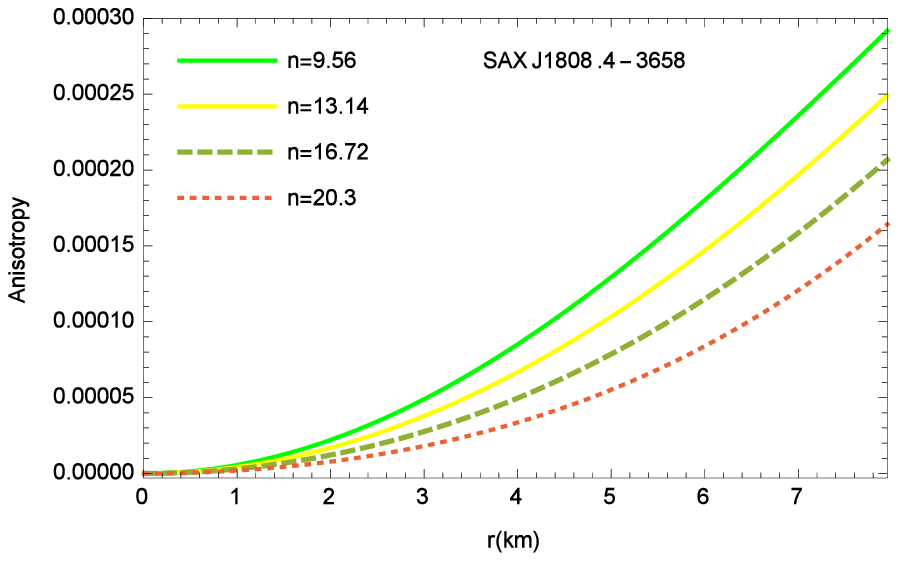}
\caption{Variation of anistropy $\Delta(r)$ with $r$ for (i)PSR J1614-2230 with mass $M=1.97 M_{\odot}$ and radius $R=9.69$km for the models $n=13.5, ~18.66, ~ 23.82, ~28.98$; (ii) SAX J1808.4-3658 with mass $M=0.9 M_{\odot}$ and radius $R=7.951$km for the models $n=~9.56,~13.14,~16.72,~20.3$ and the values of $b = 0.0001/km^2$, $c=2.5$.}
\label{delta}
\end{figure}

\twocolumn
\footnotesize{{\bf{Acknowledgments}} } The authors are thankful to the learned referee for the valuable comments and suggestions to improve the paper.


\end{document}